\documentclass{article}
\usepackage{spconf,amsmath,graphicx}

\usepackage{epsfig,psfrag}
\usepackage{pst-all}
\usepackage{amsmath,amsthm,amssymb,amsfonts,upref,cite,epsf,color,bm}
\usepackage{graphicx}
\usepackage{color}
\newcommand{\ist}{\hspace*{.2mm}}
\newcommand{\rmv}{\hspace*{-.2mm}}
\DeclareMathOperator{\supp}{supp}

\DeclareMathOperator*{\argmax}{arg\,max}

\newtheorem{theorem}{Theorem}[section]
\newcommand{\eq}{\,=\,}
\newcommand{\x}{{\bf x}}

\newcommand{\be}{\begin{equation}}
\newcommand{\ee}{\end{equation}}

\ninept

\title{Performance Bounds for Sparse Parametric Covariance Estimation\\[.5mm] 
in Gaussian Models}
\name{Alexander Jung$^{\ist 1}\rmv\rmv$, Sebastian Schmutzhard$^{\ist\ist 2}\rmv\rmv$, Franz Hlawatsch$^{\ist 1}\rmv\rmv$, 
and Alfred O. Hero III$^{\ist\ist 3}$\thanks{\hspace*{-5mm}This work 
was supported by the FWF under Grants S10602-N13 
and S10603-N13 
within the National Research Network SISE, and by the WWTF under Grant MA 07-004 (SPORTS).}\vspace{-1.5mm}
}
\address{\normalsize $^1$Institute of Telecommunications, Vienna University of Technology, Austria; \{ajung,$\ist$fhlawats\}@nt.tuwien.ac.at\\[-0.5mm]
\normalsize $^2$NuHAG, Faculty of Mathematics, University of Vienna, Austria; sebastian.schmutzhard@univie.ac.at\\[-0.5mm]
\normalsize $^3$Department of Electrical Engineering and Computer Science, University of Michigan, Ann Arbor, MI, USA; hero@eecs.umich.edu
}

\begin{document}
\maketitle

\newtheorem{definition}{Definition}[section]
\newtheorem{lemma}[theorem]{Lemma}
\newtheorem{proposition}[theorem]{Proposition}
\newtheorem{corollary}[theorem]{Corollary}

\newcommand{\CRBfull} {{C}ram\'{e}r--{R}ao bound}


\maketitle

\begin{abstract}
We consider estimation of a sparse parameter vector that determines the covariance 
matrix of a Gaussian random vector via a sparse expansion 
into known ``basis matrices.'' Using the theory of reproducing kernel Hilbert spaces, we derive lower bounds 
on the variance of estimators 
with a given mean function. This includes unbiased estimation as a special case. 
We also present a numerical comparison of our lower bounds with the variance of two standard estimators 
(hard-thresholding  estimator and maximum likelihood estimator).
\end{abstract}
\begin{keywords}Sparsity, sparse covariance estimation, variance bound, reproducing kernel Hilbert space, RKHS.

\end{keywords}


%

\vspace{-.2mm}

\section{Introduction}
\label{sec_intro} 

\vspace{-1mm}

We consider a Gaussian signal vector $\mathbf{s} \rmv \in\rmv \mathbb{R}^{M}\!$, $\mathbf{s} \rmv\sim\rmv \mathcal{N} (\bm{\mu}, \mathbf{C})$
embedded in white
Gaussian noise $\mathbf{n} \rmv\sim\rmv \mathcal{N} (\mathbf{0}, \sigma^{2} \mathbf{I})$. The observed vector \nolinebreak 
\vspace{-.5mm}
is
\begin{equation}
\label{equ_obs_model}
\mathbf{y} \ist=\ist \mathbf{s} + \mathbf{n} \,,
\vspace{-.5mm}
\end{equation}
where $\mathbf{s}$ and $\mathbf{n}$ are independent and the signal mean $\bm{\mu}$ and noise variance $\sigma^{2}$ are known.
In what follows, we assume $\bm{\mu} \!=\! \mathbf{0}$ since a nonzero $\bm{\mu}$ can always be subtracted from $\mathbf{s}$.  
The signal covariance matrix $\mathbf{C}$ is unknown; we will parameterize it according 
\vspace{-.5mm}
to
\begin{equation}
\label{equ_cov_model}
\mathbf{C} \ist=\ist \mathbf{C}(\mathbf{x}) \ist\triangleq\ist \sum_{k=1}^{N} x_{k} \mathbf{C}_k \,,
\vspace{-.5mm}
\end{equation}
with unknown nonrandom 
coefficients $x_k \!\ge\! 0$ 
and known positive semidefinite ``basis matrices'' $\mathbf{C}_k$.
Thus, estimation of the signal covariance matrix $\mathbf{C}$ reduces to estimation of the coefficient vector 
$\mathbf{x} \triangleq (x_1,\ldots,x_N)^T \!\rmv\in\rmv \mathbb{R}_{+}^{N}$.

Our central assumption 
is that $\mathbf{x}$ is \emph{$S$-sparse}, i.e., at most $S$ coefficients $x_k$ are nonzero. We can formulate this as
\begin{equation}
\label{equ_sparsity_constr}
\mathbf{x} \in \mathcal{X}_{S,+} \triangleq \big\{ \mathbf{x}' \!\rmv\in\rmv \mathbb{R}_{+}^{N} \ist\big|\ist {\| \mathbf{x}' \|}_{0} \rmv\leq\rmv S \big\} \,. 
\vspace{-.5mm}
\end{equation}
The sparsity degree $S$ is supposed known; however, 
the set of positions of the nonzero entries of $\mathbf{x}$ (denoted by ${\rm supp}(\mathbf{x})$; note that ${|{\rm supp}(\mathbf{x})| = \| \mathbf{x} \|}_{0} \le S$) 
is unknown. Typically, $S \!\ll\! N$.  
We will 
refer to 
\eqref{equ_obs_model}--\eqref{equ_sparsity_constr} as the \emph{sparse covariance model} (SCM). 
The SCM 
and estimation of 
$\mathbf{x}$ are relevant, e.g., in time-frequency (TF) analysis \cite{fla-book2,hla-book}, 
where the basis
matrices $\mathbf{C}_{k}$ correspond to disjoint TF regions 
and $x_{k}$ represents 
the mean signal power 
in the $k\ist$th TF region.
An 
application is cognitive radio scene analysis \cite{CognitiveRadio2005}. 

The problem we will study is estimation of 
$\mathbf{z} \triangleq \mathbf{g}(\mathbf{x}) \rmv \in\rmv \mathbb{R}^{K}\rmv$ from $\mathbf{y}$,
where $\mathbf{g}(\cdot)$ is a known 
function. This includes estimation of $\mathbf{x}$ and, less trivially, 
of a linear combination of the $x_{k}$. 
In the TF application mentioned above, the latter case corresponds to a linear combination of the mean signal powers in the various TF regions.

In this paper, building on 
\cite{Parzen59,Duttweiler73b}, we use the theory
of \emph{reproducing kernel Hilbert spaces} (RKHS) 
to derive lower bounds on the variance of estimators of $\mathbf{z}$. The estimators are required to have a prescribed differentiable 
mean function; this includes the case of unbiased estimation.
They are allowed to exploit the known sparsity of $\mathbf{x}$.
The RKHS framework has been previously proposed for a fundamentally different problem of sparsity-exploiting estimation in \cite{RKHSAsilomar2010}. 

Sparsity-exploiting estimation of 
$\mathbf{C}$ and of 
$\mathbf{C}^{-1}\rmv$ was considered recently 
in \cite{RueBue09} and in \cite{CovEstDecompGraphModel}, respectively. In both cases, the sparsity assumption was placed on $\mathbf{C}^{-1}\!$, which corresponds to 
a sparse graphical model for $\mathbf{s}$.
Our SCM approach \eqref{equ_cov_model}, \eqref{equ_sparsity_constr} is clearly different: while the coefficient
vector $\mathbf{x}$ is assumed sparse, the matrices $\mathbf{C}$ or $\mathbf{C}^{-1}\rmv$ need not be sparse.

This paper is organized as follows.
In Section \ref{sec_RKHS_intro}, we review minimum-variance estimation and the RKHS framework.  
In Section \ref{sec_lower_bounds}, we use RKHS theory to derive lower variance bounds for the SCM.  
The special case of unbiased estimation is considered in Section \ref{sec_unbiased_SDCM}.
Finally, Section \ref{sec_simu} presents a numerical comparison of our bounds with the variance of two established estimation schemes.

\vspace{-.7mm}

\section{RKHS formulation of minimum-variance estimation}
\label{sec_RKHS_intro} 



\subsection{Minimum-Variance Estimation}\label{sec_min_var}


The estimation error incurred by an estimator $\hat{\mathbf{z}}(\mathbf{y})$ of $\mathbf{z} \!=\! \mathbf{g}(\mathbf{x})$ can be quantified by the mean squared error (MSE)
$\varepsilon(\hat{\mathbf{z}}(\cdot), \mathbf{x}) \triangleq \mathsf{E}_{\mathbf{x}} \big\{ {\| \hat{\mathbf{z}}(\mathbf{y}) \rmv-\rmv \mathbf{z} \|}^{2}_{2} \big\}$,
where the notation $\mathsf{E}_{\mathbf{x}} \{ \cdot \}$ indicates 
that the expectation is taken with respect to the pdf $f(\mathbf{y}; \mathbf{x})$ parameterized by $\mathbf{x}$.
According to our assumptions in 
Section \ref{sec_intro},
\begin{align}
&f(\mathbf{y}; \mathbf{x}) \rmv\eq\rmv\rmv \frac{ \exp \! \big( \!-\! \frac{1}{2} \ist \mathbf{y}^{T}  \tilde{\mathbf{C}}^{-1}(\mathbf{x}) \ist\ist \mathbf{y} \big) }{ 
\big[ {(2 \pi)}^{M} \!\det \rmv\rmv \big\{ \tilde{\mathbf{C}}(\mathbf{x}) \big\} \big]^{1/2} } \,, \;\; \text{with} \;\ist
  \tilde{\mathbf{C}}(\mathbf{x}) \triangleq \mathbf{C}(\mathbf{x}) + \sigma^{2} \mathbf{I} \ist. \nonumber\\[-3mm] 
& \label{equ_param_pdf}\\[-7mm]
&\nonumber
\end{align} 

Let $z_k$ and $\hat{z}_k(\mathbf{y})$ denote the $k\ist$th entries of $\mathbf{z}$ and $\hat{\mathbf{z}}(\mathbf{y})$, respectively.
We have
$\varepsilon(\hat{\mathbf{z}}(\cdot), \mathbf{x}) = \sum_{k=1}^{N} \varepsilon(\hat{z}_{k}(\cdot),\mathbf{x})$,
where $\varepsilon(\hat{z}_{k}(\cdot),\mathbf{x})$\linebreak 
$\triangleq  \mathsf{E}_{\mathbf{x}} \big\{  [ \hat{z}_k(\mathbf{y}) - z_{k} ]^{2} \big\}$ denotes the $k\ist$th component MSE.
For our scope, minimization of 
$\varepsilon(\hat{\mathbf{z}}(\cdot), \mathbf{x})$ with respect to 
$\hat{\mathbf{z}}(\cdot)$ 
is equivalent to separate minimization of each component MSE $\varepsilon(\hat{z}_{k}(\cdot),\mathbf{x})$ with respect to 
$\hat{z}_k(\cdot)$.
We furthermore have
\begin{equation}
\label{equ_mse_bias_var}
\varepsilon(\hat{z}_{k}(\cdot),\mathbf{x}) \eq b^{2}(\hat{z}_{k}(\cdot),\mathbf{x}) \ist+\ist v(\hat{z}_{k}(\cdot),\mathbf{x}) \,,
\end{equation}
with the component bias $b(\hat{z}_{k}(\cdot),\mathbf{x}) \triangleq  \mathsf{E}_{\mathbf{x}} \{ \hat{z}_{k} (\mathbf{y}) \} - z_{k}$ and the
component variance $v(\hat{z}_{k}(\cdot),\mathbf{x}) \triangleq  \mathsf{E}_{\mathbf{x}} \big\{  [ \hat{z}_{k} (\mathbf{y}) -  \mathsf{E}_{\mathbf{x}} \{  \hat{z}_{k} (\mathbf{y})\} ]^{2} \big\}$.
A common approach to defining a ``locally optimal'' estimator $\hat{z}_{k}(\cdot)$ is to 
require $b(\hat{z}_{k}(\cdot),\mathbf{x}) =
c_{k} (\mathbf{x})$ 
for all $\mathbf{x} \!\in\! \mathcal{X}_{S,+}$, with a given bias function $c_{k}(\mathbf{x})$, and look for estimators that minimize the variance $v(\hat{z}_{k}(\cdot),\mathbf{x})$ at a 
given 
parameter vector $\mathbf{x} \!=\! \mathbf{x}_{0} \!\in\! \mathcal{X}_{S,+}$. 
It follows from \eqref{equ_mse_bias_var} that once the bias is fixed, minimizing 
$v(\hat{z}_{k}(\cdot),\mathbf{x}_0)$
is equivalent to minimizing 
$\varepsilon(\hat{z}_k(\cdot); \x_0)$.
Furthermore, fixing the bias is equivalent to fixing the mean, i.e., requiring that $\mathsf{E}_{\mathbf{x}} \{ \hat{z}_k(\mathbf{y}) \} =
\gamma_{k}(\mathbf{x})$ 
for all $\mathbf{x} \in \mathcal{X}_{S,+}$, where $\gamma_{k}(\mathbf{x}) \triangleq c_{k}(\mathbf{x}) + g_{k}(\mathbf{x})$.

In what follows, we consider a fixed component 
$k$ and drop the subscript $k$ for better readability.
Furthermore, we consider a given mean function $\gamma(\mathbf{x})$ (short for $\gamma_{k}(\mathbf{x})$) and a given nominal parameter vector $\mathbf{x}_0$.
We are interested in the minimum variance at $\mathbf{x}_0$ achievable by estimators $\hat{z}(\cdot)$ (short for $\hat{z}_{k}(\cdot)$) that have mean function 
$\gamma(\mathbf{x})$ for all $\mathbf{x} \!\in\! \mathcal{X}_{S,+}$. In order to derive a lower bound on this achievable variance,
let us consider some subset $\mathcal{D} \!\subseteq\! \mathcal{X}_{S,+}$. We denote by 
$\mathcal{B}^{\mathcal{D}}_{\gamma}(\mathbf{x}_{0})$ the set of all scalar estimators $\hat{z}(\cdot)$ whose
mean equals $\gamma(\mathbf{x})$ for all $\mathbf{x} \!\in\! \mathcal{D}$ (however, not necessarily for all $\mathbf{x} \!\in\! \mathcal{X}_{S,+}$) 
and whose variance at $\mathbf{x}_{0}$ is finite, i.e.,
\[
\mathcal{B}^{\mathcal{D}}_{\gamma}(\mathbf{x}_{0}) \,\triangleq\ist \big\{ \hat{z}(\cdot) \ist\big|\, \mathsf{E}_{\mathbf{x}}  \{ \hat{z}(\mathbf{y}) \} \rmv=\rmv Ê\gamma(\mathbf{x})  \,\, \forall \mathbf{x} \! \inÊ\! \mathcal{D} \ist,\, v(\hat{z}(\cdot), \mathbf{x}_{0}) \rmv<\rmv \infty \big\} \,.
\] 
If $\mathcal{B}^{\mathcal{D}}_{\gamma}(\mathbf{x}_{0})$ is nonempty, we consider the minimum variance 
achievable at the given parameter vector $\mathbf{x}_{0}$ by estimators $\hat{z}(\cdot) \!\in\! \mathcal{B}^{\mathcal{D}}_{\gamma}(\mathbf{x}_{0})$:
\begin{equation} 
\label{equ_min_var_constr}
L^{\mathcal{D}}_{\gamma}(\mathbf{x}_{0}) \,\triangleq\! \min_{\hat{z}(\cdot) \ist\in\ist \mathcal{B}^{\mathcal{D}}_{\gamma}(\mathbf{x}_{0})} \!v(\hat{z}(\cdot), \mathbf{x}_{0}) \,.
\vspace{-.5mm}
\end{equation} 
The use of $\min$ (rather than $\inf$) in \eqref{equ_min_var_constr} is justified by the fact that the existence of a finite minimum 
can always be guaranteed by a proper choice of $\mathcal{D}$;
a sufficient condition will be provided in Section \ref{sec_rkhs}.

Because $\mathcal{D} \!\subseteq\! \mathcal{X}_{S,+}$, $L^{\mathcal{D}}_{\gamma}(\mathbf{x}_{0})$ is a lower bound 
on the variance at $\mathbf{x}_{0}$ of any estimator $\hat{z}(\cdot)$ whose mean is
$\gamma(\mathbf{x})$ for all $\mathbf{x} \! \inÊ\! \mathcal{X}_{S,+}$ (and not just for all $\mathbf{x} \!\in\! \mathcal{D}$), i.e., 
\begin{align}
&L^{\mathcal{D}}_{\gamma}(\mathbf{x}_{0}) \ist\leq\ist v(\hat{z}(\cdot),\mathbf{x}_{0}) \,, \quad\! \text{for any $\hat{z}(\cdot)$ such that} \rule{20mm}{0mm}\nonumber\\ 
&\rule{37mm}{0mm}\mathsf{E}_{\mathbf{x}} \{ \hat{z}(\mathbf{y}) \} = \gamma(\mathbf{x}) \;\, \forall\mathbf{x} \!\in\! \mathcal{X}_{S,+} \,.
\label{equ_L_variance}
\end{align}

\vspace{-5mm}

\subsection{RKHS Formulation}\label{sec_rkhs}


An inner product of two real random variables $a \rmv=\rmv a(\mathbf{y})$, $b \rmv=\rmv b(\mathbf{y})$ can be defined as
${\langle a, b \rangle}_{\text{RV}} \triangleq \mathsf{E}_{\mathbf{x}_{0}} \rmv\{ a(\mathbf{y}) \ist b(\mathbf{y}) \}$, with induced norm
${\| a \|}_{\text{RV}} = \sqrt{ {\langle a, a \rangle}_{\text{RV}} } = \sqrt{ \mathsf{E}_{\mathbf{x}_{0}} \rmv\{ a^2(\mathbf{y}) \} }$. Note the 
dependence on $\mathbf{x}_{0}$.
One can show that 
\eqref{equ_min_var_constr} can be rewritten formally as the following constrained norm-minimization problem: 
\begin{align}
L^{\mathcal{D}}_{\gamma}(\mathbf{x}_{0}) &\eq \min_{\hat{z}(\cdot)} {\| \hat{z} \|}_{\text{RV}}^2 - \gamma^{2}(\mathbf{x}_{0}) \nonumber \\[-.5mm]  
 & \rule{10mm}{0mm}\mbox{subject to} \;\; {\langle \hat{z}, \rho_{\mathbf{x}} \rangle}_{\text{RV}} = \gamma(\mathbf{x}) \,\,\, \forall \mathbf{x} \!\in\! \mathcal{D} \, .
\label{equ_opt_norm_RV} \\[-5.5mm]
\nonumber 
\end{align} 
Furthermore, if $\mathcal{B}^{\mathcal{D}}_{\gamma}(\mathbf{x}_{0})$ is nonempty, the existence of a finite minimum in 
\eqref{equ_min_var_constr}, \eqref{equ_opt_norm_RV} can be guaranteed by choosing $\mathcal{D}$ such that \cite{Parzen59,Duttweiler73b} 
\begin{equation} 
\label{equ_finite_squared_norm}
{\| \rho_{\mathbf{x}} \|}_{\text{RV}}^2 \ist\equiv\ist\ist \mathsf{E}_{\mathbf{x}_{0}} \rmv\big\{ \rho_{\mathbf{x}}^2(\mathbf{y}) \big\} \ist<\ist \infty  
  \quad \forall \mathbf{x} \!\in\! \mathcal{D} \,,
\vspace{-2mm}
\end{equation} 
where 
\vspace{.5mm}
\begin{equation}
\label{equ_rho_def}
\rho_{\mathbf{x}}(\mathbf{y})Ê\ist\triangleq\ist \frac{f(\mathbf{y};\mathbf{x})}{f(\mathbf{y};\mathbf{x}_{0})} \,. 
\vspace{1mm}
\end{equation} 

According to \cite{Parzen59}, the solutions of 
\eqref{equ_opt_norm_RV} can 
be described 
using an RKHS $\mathcal{H}(R)$ with kernel $R(\mathbf{x}_{1}, \mathbf{x}_{2}) \!: \mathcal{D} \!\times\! \mathcal{D} \rmv\to\rmv \mathbb{R}$ given by
\begin{equation} 
\label{equ_kernel_generic}
R(\mathbf{x}_{1}, \mathbf{x}_{2}) \ist=\ist {\langle \rho_{\mathbf{x}_{1}}, \rho_{\mathbf{x}_{2}} \rangle}_{\text{RV}} 
   \ist=\ist \mathsf{E}_{\mathbf{x}_{0}} \rmv\{ \rho_{\mathbf{x}_{1}}\rmv (\mathbf{y}) \ist \rho_{\mathbf{x}_{2}} \rmv (\mathbf{y}) \} \,. 
\end{equation} 
Note that $R(\mathbf{x}_{1}, \mathbf{x}_{2})$ and $\mathcal{H}(R)$ depend on $\mathbf{x}_{0}$.
Inserting \eqref{equ_rho_def} and \eqref{equ_param_pdf} into \eqref{equ_kernel_generic} yields the expression 
\begin{align} 
R(\mathbf{x}_{1}, \mathbf{x}_{2}) &\eq\rmv \big[ \rmv\det \rmv\rmv\big\{ \tilde{\mathbf{C}}(\mathbf{x}_{0}) \big\} \big]^{1/2} \ist 
  \big[ \rmv\det \rmv\rmv \big\{ \tilde{\mathbf{C}}(\mathbf{x}_{1}) \ist \tilde{\mathbf{C}}(\mathbf{x}_{2}) \rule{17mm}{0mm} \nonumber \\
&\rule{6mm}{0mm} \cdot\rmv \big( \tilde{\mathbf{C}}^{-1}(\mathbf{x}_{1}) + \tilde{\mathbf{C}}^{-1}(\mathbf{x}_{2}) - \tilde{\mathbf{C}}^{-1}(\mathbf{x}_{0}) \big) \big\} \big]^{-1/2} , \nonumber \\[-5.2mm]
& \label{equ_kernel_SCM} 
\end{align} 
where as before $\tilde{\mathbf{C}}(\mathbf{x}) = \mathbf{C}(\mathbf{x}) + \sigma^{2} \mathbf{I}$. 
The RKHS $\mathcal{H}(R)$ is a Hilbert space of functions $f \!:\rmv \mathcal{D} \!\to\! \mathbb{R}$ that is defined as the 
closure of the linear span of the set of functions ${ \{ f_{\mathbf{x}'}(\x) = R(\x, \mathbf{x}') \} }_{\mathbf{x}' \in \mathcal{D}}$ 
This closure is taken with respect to the topology that is given by the inner product ${\langle \cdot\ist\ist , \cdot \rangle}_{\mathcal{H}(R)}$ defined via the \emph{reproducing property} \cite{aronszajn1950} 
\[
\big\langle f(\cdot) , R(\cdot,\mathbf{x}') \big\rangle_{\mathcal{H}(R)} \ist=\ist f(\mathbf{x}') \,, \quad f \rmv\rmv\in\rmv \mathcal{H}(R) \ist\ist , \; \mathbf{x}' \!\rmv \in\! \mathcal{D} \,.
\vspace{-1mm}
\]
The induced norm is ${\| f \|}_{\mathcal{H}(R)} = \sqrt{ {\langle f , f \rangle}_{\mathcal{H}(R)} }$.

It can be shown \cite{aronszajn1950, Parzen59} that if $\mathcal{D}$ satisfies \eqref{equ_finite_squared_norm}, then
$\gamma \!\in\! \mathcal{H}(R)$ is necessary and sufficient for $\mathcal{B}^{\mathcal{D}}_{\gamma}(\mathbf{x}_{0})$ to be nonempty 
and the minimum value $L^{\mathcal{D}}_{\gamma}(\mathbf{x}_{0})$ 
in \eqref{equ_min_var_constr}, \eqref{equ_opt_norm_RV} to exist and be given 
\vspace{-.5mm}
by 
\begin{equation}
\label{equ_relation_bound_squared_norm}
L^{\mathcal{D}}_{\gamma}(\mathbf{x}_{0}) \,=\, {\| \gamma \|}^{2}_{\mathcal{H}(R)} \rmv- \gamma^{2}(\mathbf{x}_{0}) \,. 
\vspace{-.7mm}
\end{equation}

\section{Lower Bounds on the Estimator Variance} 
\label{sec_lower_bounds}

\vspace{-.7mm}

According to \eqref{equ_relation_bound_squared_norm}, any lower bound on ${\| \gamma \|}^{2}_{\mathcal{H}(R)}$ 
entails a lower bound on $L_{\gamma}(\mathbf{x}_{0})$.
For mathematical tractability, we hereafter assume that the basis matrices 
$\mathbf{C}_k$ in \eqref{equ_cov_model} are projection matrices on
orthogonal subspaces of $\mathbb{R}^{M}\rmv\rmv$. Thus, they can be written as
\begin{equation}
\label{equ_matrices_spec_decom}
\mathbf{C}_k \rmv\eq\rmv \sum_{i=1}^{r_{k}} \rmv\mathbf{u}_{m_{k,i}} \rmv\mathbf{u}_{m_{k,i}}^{T} \,, \quad k = 1,\ldots,N \,,
\vspace{-1.3mm}
\end{equation}
where ${\{ \mathbf{u}_m \}}_{m=1,\ldots,M}$ is an orthonormal basis for $\mathbb{R}^{M}\rmv$ 
\vspace{-.5mm}
and the sets 
$\mathcal{U}_{k} \rmv\triangleq\rmv {\{ \mathbf{u}_{m_{k,i}} \}}_{i=1,\ldots,r_{k}}\!$
are disjoint, so that they span 
\vspace{-.3mm}
orthogonal subspaces of $\mathbb{R}^{M}\rmv\rmv$. We note that \eqref{equ_cov_model} and 
\eqref{equ_matrices_spec_decom} correspond to
a latent variable model
$\mathbf{s} = \sum_{k=1}^{N}\mathbf{s}_k$ with $\mathbf{s}_k = \sum_{i=1}^{r_{k}} \xi_{m_{k,i}} \ist \mathbf{u}_{m_{k,i}}$,
where the $\xi_{m_{k,i}}$ are independent zero-mean Gaussian with variance $x_{k}$ for all $i$, i.e., 
$\xi_{m_{k,i}} \!\rmv\sim\rmv \mathcal{N}(0, x_{k})$.
This is similar to the latent variable model used in probabilistic principal component analysis \cite{TippingProbPCA}
except that our ``factors'' $\mathbf{u}_{m}$ are fixed.
With \eqref{equ_matrices_spec_decom}, the kernel expression in \eqref{equ_kernel_SCM} simplifies 
\vspace{-3mm}
to 
\[ 
R(\mathbf{x}_{1},\mathbf{x}_{2}) \rmv\eq\rmv \frac{ \prod\limits_{k=1}^{N} (x_{0,k} \rmv\rmv+\rmv \sigma^{2})^{r_{k}} }{ \prod\limits_{k=1}^{N} 
  \!\big[ (x_{0,k} \rmv\rmv+\rmv \sigma^{2})^2 - (x_{1,k} \!-\rmv\rmv x_{0,k})(x_{2,k} \!-\rmv\rmv x_{0,k})\big]^{r_k/2} } \ist\ist, 
\vspace{-.5mm}
\] 
where, e.g., $x_{0,k}$ denotes the $k\ist$th entry of $\x_0$. We will refer to the 
SCM with basis matrices $\mathbf{C}_k$ of the form \eqref{equ_matrices_spec_decom} as the 
\emph{sparse diagonalizable covariance model} 
(SDCM).\footnote{Indeed, 
for the SDCM, the covariance matrix $\mathbf{C}(\mathbf{x})$ can be \emph{diagonalized} by a signal transformation $\mathbf{s}' \!=\rmv \mathbf{U} \mathbf{s}$,
with a unitary matrix $\mathbf{U}$ that does not depend on the true parameter vector 
$\mathbf{x}$.} 
It can be shown that, within the SDCM, a sufficient condition for \eqref{equ_finite_squared_norm}---and, thus, for the existence of a 
minimum in \eqref{equ_min_var_constr}, \eqref{equ_opt_norm_RV}---is
$x_k \rmv\rmv< 2 x_{0,k} \rmv+\rmv \sigma^{2}$ for all $k \rmv\in\rmv \{1,\ldots,N\}$. Therefore, we choose our domain as
\[ 
\mathcal{D} \eq \big\{ \mathbf{x} \!\in\! \mathcal{X}_{S,+} \ist\big| \ist\ist x_k \!< 2 x_{0,k} \rmv+\rmv \sigma^{2} \;\, \forall\, k \rmv\in\rmv \{1,\ldots,N\} \big\} \,.
\]
Note that $\mathcal{D}$ depends on $\mathbf{x}_{0}$.


We will now derive a lower bound on ${\| \gamma \|}^{2}_{\mathcal{H}(R)}$ for the SDCM. Let us assume for the moment that 
$\gamma \! \in \! \mathcal{H}(R)$. Consider 
$L$ functions $v_{l}(\x)$, $l=1, \ldots,L$,
with $v_{l} \!:\rmv \mathcal{D} \!\to\! \mathbb{R}$ and $v_{l} \!\in\! \mathcal{H}(R)$, which are orthogonal, i.e., 
${\langle v_{l}, v_{l'} \rangle}_{\mathcal{H}(R)} \!\rmv=\! 0$ if $l \!\neq\! l'\rmv$. 
Let $\mathcal{V}$ denote the subspace of $\mathcal{H}(R)$ spanned by the $v_{l}$, 
and $\mathbf{P}_{\!\mathcal{V}}$ the orthogonal projection operator on $\mathcal{V}$. 
Clearly, a lower bound on ${\| \gamma \|}^{2}_{\mathcal{H}(R)}$ is given by 
\begin{equation}
\label{equ_lower_bound_squared_norm}
{\| \mathbf{P}_{\!\mathcal{V}} \gamma \|}^{2}_{\mathcal{H}(R)} \ist\leq\ist {\| \gamma \|}^{2}_{\mathcal{H}(R)} \,. 
\end{equation}
This lower bound can be expressed as 
\begin{equation} 
\label{equ_expr_projection_0}
{\| \mathbf{P}_{\!\mathcal{V}} \gamma \|}^{2}_{\mathcal{H}(R)} \eq \sum_{l=1}^{L} \frac{ | {\langle \gamma, v_{l} \rangle}_{\mathcal{H}(R)}|^{2} }{ {\| v_l \|}^{2}_{\mathcal{H}(R)} } \,.
\end{equation}

A convenient construction of functions $v_{l}(\x)$ is  
via partial derivatives of $R(\mathbf{x}_{1},\mathbf{x}_{2})$ with respect to $\mathbf{x}_{2}$ \cite{Parzen59}.
Consider an 
index set $\mathcal{K}$ containing exactly $S$ indices from $\{1, \ldots,N\}$, i.e.,
$\mathcal{K} \rmv\subseteq\rmv \{1, \ldots,N\}$ and $| \mathcal{K} | \rmv=\rmv S$. Furthermore let $\mathbf{p}_{l} = (p_{l,1},\ldots,p_{l,N}) \rmv\in\rmv \mathbb{N}_{0}^{N}$ be
$L$ different multi-indices satisfying $\supp(\mathbf{p}_{l}) \rmv\subseteq\rmv \mathcal{K}$.
We then define
\begin{equation}
\label{equ_def_vec_part_der}
v_{l} (\mathbf{x}) \,\triangleq\ist \frac{\partial^{\mathbf{p}_{l}} R(\mathbf{x}, \mathbf{x}_{2})}{\partial \mathbf{x}_{2}^{\mathbf{p}_{l}}}  
  \bigg|_{\mathbf{x}_{2} =\ist\mathbf{x}_{0}^{\mathcal{K}}} \,, \quad\; l=1, \ldots,L \,,
\end{equation}
where $\frac{\partial^{\mathbf{p}_{l}} f(\mathbf{x})}{\partial \mathbf{x}^{\mathbf{p}_{l}}} \triangleq 
\Big( \rmv\prod_{k=1}^{N} \rmv\frac{\partial^{p_{l,k}}}{\partial x_{k}^{p_{l,k}}} \Big)  f(\mathbf{x})$ 
and $\mathbf{x}_{0}^{\mathcal{K}}$ 
is obtained from $\mathbf{x}_{0}$ 
by zeroing all entries except those whose indices are in $\mathcal{K}$.   
It can be verified that the functions $v_{l}$ 
are orthogonal, i.e., 
\begin{equation}
\label{equ_inner_prod_part}
{\langle v_{l} , v_{l'} \rangle}_{\mathcal{H}(R)} =\ist  q_{l}(\mathbf{x}_{0}) \ist \delta_{l,l'}Ê\,, 
\end{equation}
where $q_l(\mathbf{x}_{0}) =  \frac{\partial^{\mathbf{p}_{l}} \partial^{\mathbf{p}_{l}} R(\mathbf{x}_{1}, \mathbf{x}_{2})}{\partial \mathbf{x}_{1}^{\mathbf{p}_{l}} \partial \mathbf{x}_{2}^{\mathbf{p}_{l}}}  \Big|_{\mathbf{x}_{1} =\ist \mathbf{x}_{2}=\ist\mathbf{x}_{0}^{\mathcal{K}}} $. 
Furthermore \cite{Parzen59}, 
\begin{equation} 
\label{equ_inner_prod_part_eval}
{\langle f , v_{l} \rangle}_{\mathcal{H}(R)} \rmv\eq\rmv \frac{\partial^{\mathbf{p}_{l}} f(\mathbf{x})}{\partial \mathbf{x}^{\mathbf{p}_{l}}}  \bigg|_{\mathbf{x} \ist=\ist \mathbf{x}_{0}^{\mathcal{K}}}
  \quad\, \text{for any} \;\, f \!\in\! \mathcal{H}(R) \,.
\end{equation}
Using \eqref{equ_inner_prod_part} and \eqref{equ_inner_prod_part_eval} in \eqref{equ_expr_projection_0}, we obtain
\be
\label{equ_expr_projection}
{\| \mathbf{P}_{\!\mathcal{V}} \gamma \|}^{2}_{\mathcal{H}(R)} \rmv\eq\rmv \sum_{l=1}^{L} \frac{1}{q_{l}(\mathbf{x}_{0})} \Bigg| \frac{\partial^{\mathbf{p}_{l}} \gamma(\mathbf{x})}{\partial \mathbf{x}^{\mathbf{p}_{l}}} \bigg|_{\mathbf{x} \ist=\ist \mathbf{x}_{0}^{\mathcal{K}}} \Bigg|^{2}.
\ee
Finally, combining \eqref{equ_L_variance}, \eqref{equ_relation_bound_squared_norm}, \eqref{equ_lower_bound_squared_norm}, and \eqref{equ_expr_projection},
we arrive at the following bound.
(Hereafter, we again explicitly indicate the 
\vspace{1mm}
index $k$.) 

\begin{theorem} 
\label{thm_main_lower_bound}
For the SDCM, let $\hat{z}_{k}(\cdot)$ be any estimator of 
$z_k \!=\! g_{k}(\mathbf{x})$ 
whose mean equals $\gamma_{k}(\mathbf{x})$ for all $\x \rmv\in\rmv \mathcal{X}_{S,+}\,$ and whose variance 
at a fixed $\mathbf{x}_{0} \rmv\in\rmv \mathcal{X}_{S,+}$ is finite. Then, this variance 
satisfies
\begin{equation}
\label{equ_main_lower_bound}
v(\hat{z}_{k}(\cdot), \mathbf{x}_{0}) \,\geq\, \sum_{l=1}^{L} \frac{1}{q_{l}(\mathbf{x}_{0})} \Bigg| \frac{\partial^{\mathbf{p}_{l}} \gamma_{k}(\mathbf{x})}{\partial \mathbf{x}^{\mathbf{p}_{l}}}  
  \bigg|_{\mathbf{x} \ist=\ist \mathbf{x}_{0}^{\mathcal{K}}} \Bigg|^{2} \rmv-\ist \gamma_{k}^{2}(\mathbf{x}_{0}) \,,
\end{equation} 
for any choice of $L$ different 
$\mathbf{p}_{l} \rmv\in\rmv \mathbb{N}_{0}^{N}$ such that $\supp(\mathbf{p}_{l}) \rmv\subseteq\rmv \mathcal{K}$, where 
$\mathcal{K} \rmv\subseteq\rmv \{1, \ldots,N\}$ is an arbitrary set of $S$ different indices.
The lower bound \eqref{equ_main_lower_bound} is achieved by an estimator 
$\hat{z}_{k}(\cdot)$
if and only if there are nonrandom coefficients $a_{l} \!\in\rmv\rmv \mathbb{R}$ such that 
\[
\hat{z}_{k}(\mathbf{y})  \ist= \sum_{l=1}^{L}  a_{l} \ist \frac{\partial^{\mathbf{p}_{l}} \rho_{\mathbf{x}}(\mathbf{y}) }{\partial \mathbf{x}^{\mathbf{p}_{l}}}  
  \bigg|_{\mathbf{x}=\ist\mathbf{x}_{0}^{\mathcal{K}}}
\] 
with the random variables
$\rho_{\mathbf{x}}(\mathbf{y})$ 
being defined in \eqref{equ_rho_def}. 
\vspace{.3mm}
\end{theorem} 
Note that the bound in \eqref{equ_main_lower_bound} depends on $\gamma_{k}(\mathbf{x})$ only via a finite number 
of partial derivatives of $\gamma_{k}(\mathbf{x})$ at 
$\mathbf{x} \!=\! \mathbf{x}_{0}^{\mathcal{K}}$. Thus, it only depends on the local behavior of the prescribed mean or bias. 
We furthermore note that Theorem \ref{thm_main_lower_bound} 
does not mention the condition $\gamma_{k} \!\in\rmv\rmv \mathcal{H}(R)$ we used in its derivation.
This is no problem because it can be shown \cite{Parzen59} that if $\gamma_{k} \!\notin\rmv\rmv \mathcal{H}(R)$, 
there exists no estimator that has mean $\gamma_{k}(\mathbf{x})$ for all $\x \rmv\in\rmv \mathcal{X}_{S,+}$ and finite variance at $\mathbf{x}_{0}$. 

\vspace{-.7mm}


\section{Special Case: Unbiased Estimation} 
\label{sec_unbiased_SDCM}

\vspace{-.7mm}

In this section, we evaluate the bound \eqref{equ_main_lower_bound} for the important special case of unbiased estimation of $\mathbf{x}$,
i.e., for $z_k \!=\! g_k(\mathbf{x}) \!=\! x_{k}$ and $c_k(\mathbf{x}) \!\equiv\! 0$
or equivalently $\gamma_k(\mathbf{x}) \!=\! x_{k}$. To obtain a simple expression, we use $L \!=\! 2$ and particular choices of $\mathcal{K}$ and
$\mathbf{p}_{l}$ ($l \!=\! 1,2$). Specifically, using $\mathcal{K} = \{k\} \cup \mathcal{L}$, 
where 
$\mathcal{L}$ consists of the indices of the $S\!-\!1$ largest entries of the vector 
that is obtained from 
$\mathbf{x}_{0}$ 
 by zeroing the 
$k$th entry, and
$\mathbf{p}_{1} \!=\! \mathbf{0}$ and $\mathbf{p}_{2} \!=\!\mathbf{e}_k$, where $\mathbf{e}_{k}$ denotes the $k\ist$th column of the identity matrix, 
the following variance bound is obtained from 
\vspace{1mm}
Theorem \ref{thm_main_lower_bound}.

\begin{corollary}
\label{corr_SDCM_unbiased} 
For the SDCM, let $\hat{x}_{k}(\cdot)$ be any estimator of 
$x_{k}$ 
that is unbiased (i.e., $\gamma_{k}(\mathbf{x}) = x_{k}$) for all $\x \rmv\in\rmv \mathcal{X}_{S,+}\,$ and whose variance 
at a fixed $\mathbf{x}_{0} \rmv\in\rmv \mathcal{X}_{S,+}$ is finite. Then, this variance satisfies
\begin{align} 
&v(\hat{x}_{k}(\cdot), \mathbf{x}_{0}) \nonumber \\[.5mm]
& \;\; \geq \begin{cases}   
 \displaystyle \frac{2} {r_{k}}(x_{0,k} \rmv+\rmv \sigma^{2})^2 , & k \rmv\in\rmv \supp(\mathbf{x}_{0}) \\[2.5mm]Ê
 \displaystyle \frac{2} {r_{k}} \,\frac{ \sigma^{4} \ist \big[ \big( \xi(\mathbf{x}_{0}) \rmv+\rmv \sigma^{2} \big)^{2} \rmv-\ist \xi^{2}(\mathbf{x}_{0}) \big]^{r_{j_{0}}/2}}{\big( \xi(\mathbf{x}_{0}) \rmv+\rmv \sigma^{2} \big)^{r_{j_{0}}}} \,, 
   &k \rmv\not\in\rmv \supp(\mathbf{x}_{0}) \ist, \end{cases} \nonumber\\[-2mm]
& \label{equ_corr_lower_bound}\\[-7mm]
& \nonumber
\end{align}
where $\xi(\mathbf{x}_{0})$, $j_{0}$ denote the value and index respectively of the $S$-largest entry of $\mathbf{x}_{0}$. 
\vspace{.1mm}
\end{corollary}

The lower bound \eqref{equ_corr_lower_bound} 
can be achieved at least in the following 
two cases: (i) if $k \in \supp(\mathbf{x}_{0})$, and (ii) for any $k \rmv\in\rmv \{1,\ldots,N\}$ if ${\| \mathbf{x}_{0} \|}_{0} \rmv<\rmv S$ 
(note that this condition implies $\xi(\mathbf{x}_{0}) \rmv=\rmv 0$). 
In both cases, 
the estimator given 
\vspace{-1mm}
by
\begin{align} 
&\hat{x}_{k}(\mathbf{y}) \rmv\eq\rmv \beta_k(\mathbf{y}) \rmv-\rmv \sigma^{2} , \quad \text{with} \;\; 
  \beta_k(\mathbf{y}) \triangleq \frac{1}{r_{k}} \rmv\sum_{i=1}^{r_k} \rmv\big( \mathbf{u}^{T}_{m_{k,i}} \ist \mathbf{y} \big)^{2} , \nonumber\\[-3mm]
&\label{equ_tight-est}\\[-6mm]
\nonumber
\end{align} 
is unbiased and its variance achieves the bound 
\eqref{equ_corr_lower_bound}.
This estimator does not 
use the sparsity information and does not depend on 
$\x_{0}$. 

Let us define a ``signal-to-noise ratio'' (SNR) quantity as $\mbox{SNR} \triangleq \xi(\mathbf{x}_{0})/\sigma^{2}\rmv\rmv$.
For $\mbox{SNR}(\mathbf{x}_{0}) \!\ll\! 1$, the lower bound \eqref{equ_corr_lower_bound} is approximately
$ \frac{2} {r_{k}}(x_{0,k} \rmv+\rmv \sigma^{2})^2$ for any $k$, which does not depend on
$S$ and moreover equals the variance of the 
unbiased estimator \eqref{equ_tight-est}. 
Since that estimator does not exploit any sparsity information, Corollary \ref{corr_SDCM_unbiased} suggests that, in the low-SNR regime, 
unbiased estimators cannot exploit the prior information that $\x$ is $S$-sparse. 
However, in the high-SNR regime ($\mbox{SNR}(\mathbf{x}_{0}) \!\rightarrow\! \infty$), \eqref{equ_corr_lower_bound} becomes
$\frac{2} {r_{k}}(x_{0,k} \rmv+\rmv \sigma^{2})^2$ for $k \rmv\in\rmv \supp(\mathbf{x}_{0})$ and $0$ for $k \rmv\not\in\rmv \supp(\mathbf{x}_{0})$, which can be shown to equal
the variance of the oracle estimator that knows 
$\supp(\mathbf{x}_{0})$ (this oracle estimator yields $\hat{x}_{k} \!=\! x_{0,k} \!=\! 0$ 
for all $k \rmv\not\in\rmv \supp(\mathbf{x}_{0})$). 
The transition of the lower bound \eqref{equ_corr_lower_bound} from the low-SNR regime to the high-SNR regime 
has a polynomial characteristic; it is thus much slower than the exponential transition of an analogous 
lower bound recently
derived in \cite{RKHSAsilomar2010} for the \emph{sparse linear model}.
This slow
transition suggests that the optimal estimator for low SNR---which ignores the sparsity information---will also be nearly optimal over a relatively wide SNR range. 
This further suggests that, for covariance estimation based on the SDCM, 
prior information of sparsity is not as helpful as for estimating the mean of a Gaussian random vector
based on the sparse linear model \cite{RKHSAsilomar2010}.

In the special case where $S\!=\!1$ and $\mathbf{x}_{0} \!\neq\! \mathbf{0}$, let $\xi_{0}$ and $j_{0}$ denote, respectively, the value and index 
of the single nonzero entry of 
$\mathbf{x}_{0} \rmv\in \mathcal{X}_{1,+}$. 
Consider the estimator $\hat{\mathbf{x}}^{(\mathbf{x}_{0})}(\cdot)$ given componentwise by 
\begin{equation}
\hat{x}^{(\mathbf{x}_{0})}_{k}(\mathbf{y}) = \begin{cases} 
  \beta_k(\mathbf{y}) \rmv-\rmv \sigma^{2} , & k = j_{0} \\[.5mm]
  \alpha(\mathbf{y};\mathbf{x}_{0}) \ist \big( \beta_k(\mathbf{y}) \rmv-\rmv \sigma^{2} \big) \,, & k \neq j_{0} \,,  \end{cases}
\end{equation}
where $\alpha(\mathbf{y};\mathbf{x}_{0}) \triangleq a(\mathbf{x}_{0}) \exp\rmv\rmv\big(\!\rmv-\rmv\rmv r_{j_{0}} b(\mathbf{x}_{0}) \ist \beta_{j_{0}}(\mathbf{y})\big)$
\vspace*{-.7mm}
with $a(\mathbf{x}_{0}) \triangleq \frac{[ (\xi_{0}+\sigma^{2})^2 - \xi_{0}^{2}]^{r_{j_{0}}/2}}{\sigma^{r_{j_{0}}} (\xi_{0} + \sigma^{2})^{r_{j_{0}}/2}}$ 
and $b(\mathbf{x}_{0}) \triangleq \frac{1}{2}\ist\big(\frac{1}{\sigma^{2}} - \frac{1}{\xi_{0} + \sigma^{2}} \big)$. 
One can show
using RKHS theory that $\hat{\mathbf{x}}^{(\mathbf{x}_{0})}(\cdot)$ is unbiased and has the minimum variance 
achievable by unbiased estimators at any 
$\mathbf{x}_{0} \in \mathcal{X}_{1,+}$ with $\mathbf{x}_{0} \!\neq\! \mathbf{0}$. 
Note that this estimator depends explicitly on the assumed 
$\mathbf{x}_{0}$, at which it achieves minimum variance; its performance may be poor when the
true parameter vector $\mathbf{x}$ is different from $\mathbf{x}_{0}$.

\vspace{-.7mm}


\section{Numerical Results} 
\label{sec_simu}

\vspace{-.7mm}

We compare the lower bound \eqref{equ_main_lower_bound} for $\mathbf{g}(\mathbf{x}) \!=\! \mathbf{x}$ with the variance of 
two standard estimators.
The first is an ad-hoc adaptation of the hard-thresholding (HT) estimator \cite{Donoho94idealspatial} to SDCM-based covariance estimation.
It is defined componentwise as (cf.\ \eqref{equ_tight-est})
\[
\hat{x}_{k,\text{HT}}(\mathbf{y}) \ist\triangleq\ist \frac{1}{r_k} \rmv\sum_{i=1}^{r_k} \varphi_\tau^2 \big( \mathbf{u}^{T}_{m_{k,i}} \ist \mathbf{y} \big)  - \sigma^{2} ,
\]
where $ \varphi_\tau \!:\rmv \mathbb{R} \!\to\! \mathbb{R}$ denotes the hard-thresholding function with threshold $\tau \rmv\ge\rmv 0$, 
i.e., $\varphi_\tau(y)$ is $y$ for $|y| \geq \tau$ and $0$
else.
The second standard method 
is the maximum likelihood (ML) 
estimator  
\[ 
\hat{\mathbf{x}}_{\text{ML}}(\mathbf{y}) \,\triangleq\, \argmax_{\mathbf{x}' \in \mathcal{X}_{S,+}} \ist f ( \mathbf{y};  \mathbf{x}' ) \,.
\vspace{-1mm}
\] 
For the SDCM, 
one can 
show that
\[ 
\hat{x}_{k,\text{ML}}(\mathbf{y}) \eq \begin{cases} 
  \beta_k(\mathbf{y}) \rmv-\rmv \sigma^{2} , & k \rmv\in\rmv \mathcal{L}_1 \!\cap\rmv \mathcal{L}_2 \\ 
  0 \,, &\mbox{else} \ist , \end{cases}
\vspace{-.5mm}
\]
where 
$\mathcal{L}_1$ consists of the $S$ indices $k$ for which 
$r_{k} \big[ \beta_{k}(\mathbf{y})/\sigma^2 \! -Ê\! \ln \ist (\beta_{k}(\mathbf{y})/\sigma^{2}) \rmv-\! 1 \big]$ (with $\ln \rmv=\rmv \log_e$)
is largest, and 
$\mathcal{L}_2$ consists of all indices $k$ for which $\beta_k(\mathbf{y}) \rmv\geq\rmv \sigma^{2}\rmv$.

\begin{figure}
\vspace{-1mm}
\centering
\psfrag{SNR}[c][c][.9]{\uput{3.4mm}[270]{0}{\hspace{3mm}SNR [dB]}}
\psfrag{title}[c][c][.9]{\uput{2.5mm}[270]{0}{}}
\psfrag{x_0_01}[c][c][.9]{\uput{0.3mm}[270]{0}{$-20$}}
\psfrag{x_0_1}[c][c][.9]{\uput{0.3mm}[270]{0}{$-10$}}
\psfrag{x_1}[c][c][.9]{\uput{0.3mm}[270]{0}{$0$}}
\psfrag{x_10}[c][c][.9]{\uput{0.3mm}[270]{0}{$10$}}
\psfrag{x_100}[c][c][.9]{\uput{0.3mm}[270]{0}{$20$}}
\psfrag{x_1000}[c][c][.9]{\uput{0.3mm}[270]{0}{$30$}}
\psfrag{y_0}[c][c][.9]{\uput{0.1mm}[180]{0}{$0$}}
\psfrag{y_1}[c][c][.9]{\uput{0.1mm}[180]{0}{$1$}}
\psfrag{y_2}[c][c][.9]{\uput{0.1mm}[180]{0}{$2$}}
\psfrag{y_3}[c][c][.9]{\uput{0.1mm}[180]{0}{$3$}}
\psfrag{varbound}[c][c][.9]{\uput{1mm}[90]{0}{variance/bound}}
\psfrag{data2}[l][l][0.8]{bound on $v(\hat{\mathbf{x}}_{\text{ML}}(\cdot);\mathbf{x}_0)$}
\psfrag{data1}[l][l][0.8]{$v(\hat{\mathbf{x}}_{\text{ML}}(\cdot);\mathbf{x}_0)$}
\psfrag{data3}[l][l][0.8]{$v(\hat{\mathbf{x}}_{\text{HT}}(\cdot); \mathbf{x}_0)$, $\tau \!=\! 3$}
\psfrag{data5}[l][l][0.8]{$v(\hat{\mathbf{x}}_{\text{HT}}(\cdot); \mathbf{x}_0)$, $\tau \!=\! 6$}
\psfrag{data7}[l][l][0.8]{$v(\hat{\mathbf{x}}_{\text{HT}}(\cdot); \mathbf{x}_0)$, $\tau \!=\! 9$}
\psfrag{data4}[l][l][0.8]{bound on $v(\hat{\mathbf{x}}_{\text{HT}}(\cdot); \mathbf{x}_0)$, $\tau \!=\! 3$}
\psfrag{data6}[l][l][0.8]{bound on $v(\hat{\mathbf{x}}_{\text{HT}}(\cdot); \mathbf{x}_0)$, $\tau \!=\! 6$}
\psfrag{data8}[l][l][0.8]{bound on $v(\hat{\mathbf{x}}_{\text{HT}}(\cdot); \mathbf{x}_0)$, $\tau \!=\! 9$}
\psfrag{ML}[l][l][0.8]{ML}
\psfrag{T3}[l][l][0.8]{\uput{0mm}[180]{0}{HT   \!($\tau \!=\! 3$)}}
\psfrag{T6}[l][l][0.8]{HT \!($\tau \!=\! 6$)}
\psfrag{T9}[l][l][0.8]{HT \!($\tau \!=\! 9$)}
\centering
\hspace*{1.5mm}\includegraphics[height=5.4cm,width=8.7cm]{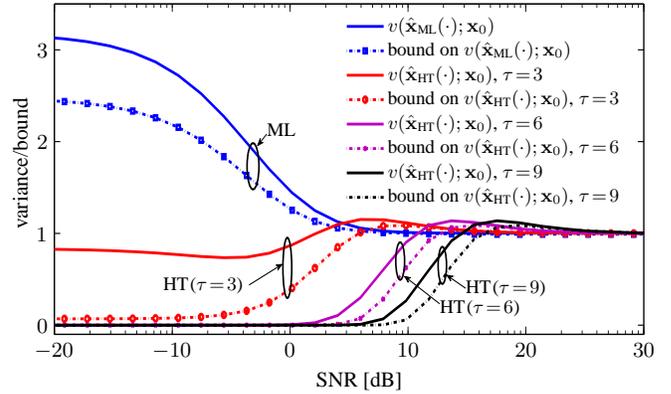}
\vspace{-3.5mm}
  \caption{Normalized variance of the HT and ML estimators and corresponding lower bounds 
  versus $\mbox{SNR} \!=\rmv\rmv \xi(\mathbf{x}_{0})/\sigma^{2}\!$, for the SDCM 
  with $N\!=\!5$, $S\!=\!1$, $\sigma^{2} \!=\!1$, and $\mathbf{C}_{k} \!=\rmv \mathbf{e}_{k} \mathbf{e}_{k}^{T}$. } 
\label{fig_bounds_1}
\vspace*{-1.5mm}
\end{figure}

For a numerical evaluation, we considered the SDCM  with  $N\!\rmv=\!5$, $S\!=\!1$, $\sigma^{2} \!=\!1$, and $\mathbf{C}_{k} \!=\rmv \mathbf{e}_{k}\mathbf{e}_{k}^{T}$.
We generated parameter vectors $\mathbf{x}_0$ with $j_0 \!=\! 1$ and different $\xi_{0}$.
In Fig.\ \ref{fig_bounds_1}, we show the 
variance at $\mathbf{x}_{0}$, 
$v(\hat{\mathbf{x}}(\cdot), \mathbf{x}_0) \!=\! \sum_{k=1}^{N} v(\hat{x}_{k}(\cdot),\mathbf{x}_0)$ 
(computed by means of numerical integration), for the HT estimator using various choices of 
$\tau$ and for the ML estimator. 
The variance is plotted versus $\mbox{SNR} = \xi(\mathbf{x}_{0})/\sigma^{2}\rmv\rmv = \xi_0/\sigma^{2}\rmv\rmv$. 
Along with each variance curve, we display a corresponding lower 
bound that was calculated by evaluating \eqref{equ_main_lower_bound} 
for each 
$k$, using for 
$\gamma_{k}(\mathbf{x})$ the mean function of the respective estimator (HT or ML), and summing over all $k$. 
(The mean functions of the HT and ML estimators were computed by means of numerical integration.)
In evaluating \eqref{equ_main_lower_bound}, we used partial derivatives of order at most $1$ in \eqref{equ_def_vec_part_der}, and
we chose for the evaluation of the lower bound $L \!=\! 2$, $\mathcal{K} \!=\rmv \{k\}$,  
$\mathbf{p}_{1} \!=\! \mathbf{0}$, and $\mathbf{p}_{2} \!=\rmv \mathbf{e}_k$. 
In Fig.\ \ref{fig_bounds_1}, all variances and bounds are normalized by $2 \ist (\xi_{0} \rmv+\rmv \sigma^{2})^{2}$, which is the variance of the
oracle estimator knowing $j_0$.

It can be seen from Fig.\ \ref{fig_bounds_1} that in the high-SNR regime, for both estimators, the gap between the variance and the corresponding lower bound 
is quite small. This indicates that the performance of both estimators is nearly optimal. However, in the low-SNR regime, 
the variances of the 
estimators tend to be
significantly higher than the 
bounds. This means that there \emph{may} be estimators 
with the same bias and mean function as that of the HT or ML estimator but a lower variance. However, the actual existence of such estimators 
is not shown by
our analysis.

\vspace{-.7mm}


\section{Conclusion} 

\vspace{-.7mm}

We considered estimation of (a function of) a sparse 
vector $\mathbf{x}$ that determines the covariance matrix of a Gaussian random vector
via a parametric covariance model.
Using RKHS theory,
we derived lower bounds on the estimator variance for a prescribed bias and 
mean function.
For the important special case of unbiased estimators of $\mathbf{x}$, we found that the transition of our bounds from low 
to high SNR
is significantly slower than that of analogous 
bounds 
for the sparse linear model 
\cite{RKHSAsilomar2010}. 
This suggests that 
the prior information of sparsity is not as helpful as for the sparse linear model. 
Numerical results showed that for low SNR, the variance of two standard estimators 
(hard-thresholding  estimator and maximum likelihood estimator) is significantly higher than our bounds. 
Hence, there might exist estimators that have the same bias and mean function as these standard estimators 
but a smaller variance. 

\renewcommand{\baselinestretch}{0.9}\normalsize\footnotesize

\bibliographystyle{IEEEtran}
\bibliography{/Users/ajung/Studium/Diss/PHDThesis/LitAJPHD.bib}

\end{document}